\documentclass[review]{elsarticle}

\PassOptionsToPackage{hyphens}{url}
\usepackage{lineno}
\usepackage{hyperref}
\usepackage{graphicx}
\usepackage{amssymb}
\usepackage[all]{xy}
\newdir{ >}{{}*!/-6pt/@{>}}
\usepackage[defblank]{paralist}
\usepackage{eurosym}
\usepackage[utf8]{inputenc}
\usepackage{enumitem}
\usepackage{caption} 
\usepackage{subcaption} 
\usepackage{stmaryrd}
\usepackage{amsmath}
\usepackage{listings}
\usepackage{comment}
\usepackage[dvipsnames]{xcolor}
\modulolinenumbers[5]

\colorlet{punct}{red!60!black}
\definecolor{background}{HTML}{EEEEEE}
\definecolor{delim}{RGB}{20,105,176}
\colorlet{numb}{magenta!60!black}
\lstdefinelanguage{json}{
	basicstyle=\ttfamily\tiny,
	numbers=left,
	numberstyle=\ttfamily\tiny,
	stepnumber=1,
	numbersep=8pt,
	showstringspaces=false,
	breaklines=true,
	frame=lines,
	backgroundcolor=\color{background},
	literate=
	*{0}{{{\color{numb}0}}}{1}
	{1}{{{\color{numb}1}}}{1}
	{2}{{{\color{numb}2}}}{1}
	{3}{{{\color{numb}3}}}{1}
	{4}{{{\color{numb}4}}}{1}
	{5}{{{\color{numb}5}}}{1}
	{6}{{{\color{numb}6}}}{1}
	{7}{{{\color{numb}7}}}{1}
	{8}{{{\color{numb}8}}}{1}
	{9}{{{\color{numb}9}}}{1}
	{:}{{{\color{punct}{:}}}}{1}
	{,}{{{\color{punct}{,}}}}{1}
	{\{}{{{\color{delim}{\{}}}}{1}
	{\}}{{{\color{delim}{\}}}}}{1}
	{[}{{{\color{delim}{[}}}}{1}
	{]}{{{\color{delim}{]}}}}{1},
}
\newcommand{\BOM}[1]{\ensuremath{\textit{BOM}_{#1}}} 
\newcommand{\fixc}[1]{\ensuremath{\textit{cf}_{#1}}} 
\newcommand{\varc}[1]{\ensuremath{\textit{cv}_{#1}}} 
\newcommand{\tspan}[1]{\ensuremath{\textit{tp}_{#1}}} 
\newcommand{\iquota}[1]{\ensuremath{\textit{g}_{#1}}} 
\newcommand{\cost}[1]{\ensuremath{\textit{c}_{#1}}} 
\newcommand{\mincost}[1]{\ensuremath{\textit{c}^{\textit{min}}_{#1}}} 
\newcommand{\scost}[1]{\ensuremath{\textit{c}^{\textit{s}}_{#1}}} 
\newcommand{\itoti}[1]{\ensuremath{\textit{PL}_{#1}}} 
\def\itot {\ensuremath{\textit{PL}}} 
\def\demand {\ensuremath{d}} 
\def\price {\ensuremath{p}} 
\newcommand\ishare[1]{\ensuremath{\textit{PL}_{#1}}} 
\def\resmax {\ensuremath{\textit{ress}}} 
\def\lvlmax {\ensuremath{\textit{levs}}} 
\def\splmax {\ensuremath{\textit{sups}}} 
\def\alcost {\ensuremath{\textit{cAll}}} 
\newcommand\qty[1] {\ensuremath{\textit{q}_{#1}}} 
\newcommand\alcosti[1]{\ensuremath{\textit{cAll}_{#1}}} 
\newcommand\coeff[1]{\ensuremath{\textit{f}_{#1}}} 
\def\SCS {\ensuremath{\textit{SCS}}} 
\def\PC {\ensuremath{\textit{PC}}} 
\def\GI {\ensuremath{\textit{R}}} 
\newcommand\GIs[1] {\ensuremath{\textit{PC}_{#1}}} 

\journal{Blockchain: Research and Application}

\begin{document}

\begin{frontmatter}

\title{Variants in managing supply chains on distributed ledgers}

\author{Paolo Bottoni, Claudio Di Ciccio}
\address{Department of Computer Science -- Sapienza University of Rome, Italy}
\author{Remo Pareschi} 
\address{Stake Lab -- University of Molise, Italy}
\author{Domenico Tortola}
\address{Department of Bioscience  and  Territory -- University of Molise, Italy}
\author{Nicola Gessa, Gilda Massa}
\address{TERIN-SEN-CROSS, ENEA, Portici (NA), Italy}

\begin{abstract}
Smart contracts show a high potential for ensuring that Supply Chain Management strategies make a qualitative leap toward higher levels of optimality, not only in terms of efficiency and profitability but also in the aggregation of skills aimed at creating the best products and services to bring to the market. In this article, we illustrate an architecture that 
employs smart contracts 
to implement various algorithmic versions of the Income Sharing principle between companies participating in a supply chain. We implement our approach on 
Hyperledger Fabric, the most widespread 
platform for private and consortium distributed ledgers, 
and discuss its suitability to our purposes by comparing this design choice with the alternative given by public blockchains, with particular attention to Ethereum.
\end{abstract}

\end{frontmatter}


\section{Introduction}\label{sec:intro}
%
%

The advent of blockchains and distributed ledgers (DLs) has brought to the fore, in addition to cryptocurrencies, highly innovative business models such as Decentralized Autonomous Organizations (DAOs) and Decentralized Finance (DeFi). 
These models were designed in the first place for virtual companies but can also be profitably translated into the ongoing digital transformation of the traditional economy, so as to contribute to the implementation of programs such as Industry 4.0. 
For this to happen, they must be applied to business processes inherent in brick-and-mortar companies. 
Supply Chain Management 
is from this point of view a domain of particular interest, by providing, on the one hand, the basis for decentralized business ecosystems compatible with the DAO model and, on the other hand, an essential component in the management of the physical goods that underlie the real economy.

In this article we intend to contribute to this evolution with a general supply chain model based on the principle of \emph{Income Sharing} (IS), according to which several companies join forces, for a specific process or project, as if they were a single company. 
Therefore, given these premises, the income is divided per a previously agreed upon distribution scheme. 
This approach is in itself more performing and effective than traditional wholesale agreements, which lack coordination among the participants in the supply chain. 
Furthermore, today it is all the more practicable by virtue of the Internet-economy and the consequent availability of platforms such as Amazon, Alibaba, Ozon and e-Bay, which small and medium-sized enterprises (SMEs) can partner with to convey their offers on markets that were beyond reach before the Internet. 
These platforms are in fact particularly effective at endowing SMEs with e-commerce, marketing and logistics functionalities that are essential for competing globally.

In implementing the IS model, various options for distributing income must be taken into account, partly due to the role played by the Internet platforms themselves in supporting the supply chains. 
At one end there is the well-known and studied \emph{Revenue Sharing} (RS)~\cite{CL05}, according to which  each of the participants in the supply chain, platform included, bears its own costs and obtains a proportional return in the form of sales revenues. 
However, other feasible options are those in which part of the costs of the participants is borne by the originator of the supply chain, or by the platform as regards its own services. 
These costs are then deducted from the distribution of the proceeds, so that it seems appropriate to speak in this case of \emph{Profit Sharing} (PS). 

As shown in~\cite{Gong2018RevenueSO}, criteria exist for quantifying, on the basis of volumes of goods handled, number of participants, and stages of the supply chain, the preferability of the various options for the supply chain stakeholders according to their roles. 
We can assume that the option to pursue is evaluated and negotiated on a project-by-project basis. 
This makes it therefore appropriate and desirable that the full range of options be made available in an IS implementation.
In addition, it is not always easy to apply the management layer of the selected IS option to supply chains of SMEs, where trust and transparency are often lacking. 
In many cases, Internet platforms 
%
%
act as arbitrators in proposing 
%
%
IS 
options to SMEs, and take on the management of the execution of the agreed upon distribution schemes. 
By doing so, however, they are in a position to impose solutions to their sole advantage as they take part in the income distribution themselves.

By virtue of their deployment on blockchains or DLs, smart contracts may well be a panacea for these issues. In fact, all that is needed is computing the costs that must initially be incurred by the participating companies and then distributing income. These tasks can be conveniently automated, provided that all stakeholders have visibility and consequent trust in the algorithm that performs them, and Internet platforms would therefore be exempted from carrying them out, while continuing to provide those services that have so effectively contributed to widening the range of commercial action of SMEs. Indeed, by making the management algorithm operate as a smart contract, we have a stringent and satisfactory response to these requirements: algorithmic automation substantially mitigates the costs and complications deriving from an additional level of human management that would otherwise be necessary, and transparency resulting from deployment on  blockchain or DL ensures the trustability of such automated management for the participants.

Bearing in mind that IS is a general scheme which groups the more specific options envisaged by RS and PS, it follows that optimal IS engineering should thus rely on a layer of fundamental functionalities to be subsequently specialised according to the choice made. To this end, we design a modular architecture 
leveraging an algorithmic basis that encompasses the viable options.
We illustrate an implementation based on Hyperledger Fabric and discuss its feasibility as a platform of choice as opposed to the Ethereum public blockchain. 

The rest of this paper is organised as follows.
After providing an extensive background in Section~\ref{sec:background}, we present a motivating scenario and general information on our approach in Section~\ref{sec:supplyChain}.
Then, Section~\ref{sec:modular} proposes a model for flexible management of smart contracts deriving from selecting specific options relative to financing the supply chain and sharing income. 
Details about the specific calculations of income sharing are given in Section~\ref{sec:algorithm}, while Section~\ref{sec:interacting} discusses the processes by which participants cooperatively introduce the information needed to run the selected algorithm, illustrating them with examples of concrete interactions. 
Finally, Section~\ref{sec:discussion} discusses implementation choices and trade-offs, and Section~\ref{sec:concls} concludes the paper.


\section{Background}\label{sec:background}
The video rental market was an incubator for the first uses of Revenue Sharing in the late 1990s, through the efforts of Blockbuster, then the industry leader, to exploit it in order to stimulate a steady growth in revenues, as described by Dana and Spier in~\cite{dana}. 
It was then systematically investigated in the seminal article by Cachon and Lariviere~\cite{CL05}. 
An early study involving an Internet platform in the set-up and run of an RS supply chain is provided by Wang~\textit{et~al}.~\cite{wang}, aiming to investigate the effect of RS on the performance of a sales channel where a supplying company uses Amazon for e-retail and logistics, thus finding out that performance, both of the overall channel and of the individual firm, depend on demand price elasticity as well as on the retailer's share of the channel cost. 
Qian~\textit{et~al}.~\cite{qian} provide a case study in the Chinese dairy sector, where a number of structural problems, including an unbalanced allocation of profits along the supply chain in favour of retailers (supermarkets) and to the disadvantage of farmers and producers, have been addressed by applying the influential three-stage RS model by Giannoccaro~and~Potrandolfo~\cite{GIANNOCCARO2004131}, with an increase in overall profitability of 12.49\%. 
The algorithmic RS methodology provided by Tononi~\textit{et~al}.~\cite{Tononi2007LaSC} is the cornerstone of the architecture described here, mainly because it lends itself easily to implementation and is at the same time highly modular and flexible so that variations such as Profit Sharing can be straightforwardly integrated, as will be detailed in Section~\ref{sec:algorithm}. 
Furthermore, it explicitly addresses the problem of trust, so as to give the best results if the participants of the supply chain are discouraged from showing production costs higher than the real ones and none of them enjoys economic and informative privileges over the others. 
These characteristics make blockchains and DLs a perfect match for its deployment. 
In the agricultural sector, the effective application of this methodology, albeit with the dis-optimisations deriving from the manual management of the coordination of the participants in the supply chain, was among the results of the LEMURE (Logistica intEgrata MUltiagente per REti di PMI)\footnote{Funded by the Italian Ministry of Research under Law 297/99, Grant agreement  2007/32458.} project. Focused on the tomato processing chain, it enjoyed improvements of up to 17\% in overall profitability. Its optimised version in the form of a smart contract operating on blockchain/DL technologies is one of the objectives of the project WEBEST (Wine EVOO Blockchain Et Smart ContracT)\footnote{Funded by the Italian Ministry of Research under the PRIN program: Research Projects of Relevant National Interest -- Call 2020.}, both projects being funded by the Italian Ministry of Research.

On the Profit Sharing side, Çanakoğlu and Bilgic~\cite{CANAKOGLU2007995} analyse the performance over multiple periods of a two-stage telecommunications supply chain, consisting of an operator and a vendor and, for optimisation purposes, suggest a PS contract in which companies share both revenues and operating costs. 
Wei~and~Choi~\cite{WEI2010255} illustrate an industrial practice of PS in the apparel sector, on the basis of which they explore the use of a wholesale pricing and profit sharing scheme for coordination of the supply chain according to the criteria of mean variance. Both of these contributions are part of the background used by Gong~\textit{et~al}.~\cite{Gong2018RevenueSO} to define selection criteria between RS and PS depending on the characteristics of the supply chain and the role of the participants.

From an organisational standpoint, the reference model is the Decentralized Autonomous Organization (DAO), widely known and discussed within the blockchain community and with a significant case history of implementations. The brainchild of the  founder  of  Ethereum  Vitalik  Buterin,%
\footnote{\url{https://blog.ethereum.org/2014/05/06/daos-dacs-das-and-more-an-incomplete-terminology-guide/}} a DAO is  an  entity that  lives  on  the  Internet  and  exists  autonomously,  relying  on  individuals  to carry out tasks aimed at the realisation of a project or the provision of a service and on algorithms to coordinate them.  This definition fits perfectly to Income Sharing, in the spectrum of variations spanning RS and PS, to the point that we could consider DAO and IS, in its deployment on blockchain or DL, as a technological case of convergent evolution.  This deployment actually extends the DAO organisational model from subjects operating in the virtual world to companies rooted in the real economy, with all the concreteness and focus that follow. This is directly reflected in the procedural characteristics of supply chains, focused as they are on bringing a product or service to the market. Therefore, participants in an IS supply chain, while choosing to join freely, once on board have every reason to collaborate closely in order to get results as quickly and profitably as possible. By contrast, an early and most ambitious DAO project, eponymously named ``The DAO''~\cite{Ban16}, was an investor-led, purely virtual venture capital fund launched on the Ethereum blockchain in April 2016, only to be disabled within a few months, due to a security attack known as the ``DAO Exploit''~\cite{Dhillon.etal/2017:TheDAOHacked}. As argued by Morrison~\textit{et~al}.~\cite{DBLP:journals/fbloc/MorrisonMW20}, lack of focus, as well as of collaboration and interaction between participants, largely explains the untimeliness in reacting effectively to the threat posed by the Exploit, which led to the subtraction of approximately \$ 60 million and to a subsequent hard fork of the Ethereum blockchain in an effort to run late for cover.
In the longer run, this incident has also resulted in a temporary slowdown in DAO projects. 
%

There is abundant literature, as well as a considerable number of ongoing projects, on the use of blockchain and distributed ledger technologies to support supply chains and collaborative processes~\cite{DBLP:journals/insk/CiccioCDGLLMPTW19,DBLP:conf/bpm/WeberXRGPM16,DBLP:conf/sac/CorradiniMMPRT20,Madsen.etal/FAB2018}, but relatively little efforts have been devoted to their use for the purpose of innovation in the way of doing business. 
Indeed, the bulk of these contributions were so far directed towards using blockchains and distributed ledgers in order to notarise the production steps along the supply chain, so as to be able to document the quality standards and compliance with current regulations of products and services.  However, there are a few theoretical contributions that are in line with our approach, which aims at exploiting these technologies in order to create an IT support for a radical business innovation. To begin with, Korpela~\textit{et~al}.~\cite{KorpelaHD17} take a Transaction Cost Economics (TCE)~\cite{coase1937nature,williamson1985economic} point  of  view  to  provide,  on  the  basis  of  empirical  evidence obtained  from  interviewing  firms  and  business  managers,  an overview  of  the  perspectives  opened  by  blockchain technology for supply chain management. 
Treiblmaier~\cite{Treiblmaier18} widens this perspective to include the points of view of
%
%
Positive Agency Theory, Resource-based View Theory and Network Theory. 
Bottoni~\textit{et~al}. engage in~\cite{BGMP+20} in an extensive discussion on the congruence between the methodological indications deriving from TCE and the blockchain-based IT architecture for the deployment of Revenue Sharing illustrated therein, which will be expanded here to fit Profit Sharing as well as Internet platforms such as Amazon and its likes to the extent that they contribute to realising the general principle of Income Sharing. As TCE favours the  keeping  in  the  firm  of  just  core  competencies  and  externalizing  all  else, the fundamental reasons for this congruence are in the ability to easily create global supply chains that promote the core competencies of companies and outsource all secondary ones by virtue of blockchain-managed trust. Alliances based on effective added value are hence enabled even between partners hitherto completely unknown to each other.  This frees companies from the yoke of local partnerships, justified by the old principle that trust is built by knowing each other and doing business together over the years, so as to be able to seek out the best partners globally. The way is thus opened to truly ``glocal'' excellence where the best, production-wise or service-wise, of a given territory is combined with the best of another area  placed at an arbitrary geographical and cultural distance. 
This congruence is all the more stringent in the context of the architectural expansion implemented here, as Internet platforms become part of the picture, with their ability to outsource and optimise services such as logistics, marketing and e-commerce. Our architecture is therefore configured as a higher-level platform, which guarantees trust in the interaction between companies and the ever more pervasive Internet platforms, so as to give the former the opportunity to take advantage of the latter without having to bend to their bargaining power.

Morrison~\textit{et~al}.~\cite{DBLP:journals/fbloc/MorrisonMW20} explore, from an Agency Theory point of view, the corporate governance implications of the DAO in its original formulation and identify possible vulnerabilities deriving from a simplistic application of algorithmic trust for its implementation. 
They address these weaknesses by proposing elements of hierarchisation and focus for DAOs 
%
%
along lines fitting our architecture.

Finally, we remark that a number of research works~\cite{Choi20,Hayrutdinov,Liu,Korsten} provide analytical models demonstrating that RS as a supply chain management methodology is optimally transferable to blockchain and smart contracts. 
They can be seen as providing formal background, albeit independently developed, to the implementations illustrated here and in~\cite{BGMP+20}.

\section{A view on supply chain management}\label{sec:supplyChain}
\begin{figure}[htb]
\centering
	\includegraphics[width=0.75\linewidth]{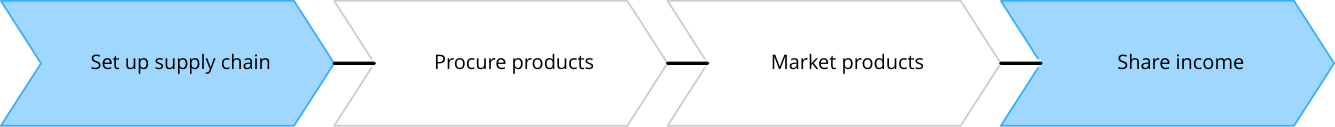}
	\caption{The process architecture at large}
	\label{fig:valuechain}
\end{figure}

Figure~\ref{fig:valuechain} depicts a map of the processes involved in the supply chain we focus on in this work~\cite{DBLP:books/sp/DumasRMR18}. 
In particular, we remark that our approach directly involves the first and the last stage in the chain (coloured in blue in the figure):
\begin{inparaenum}[(1)]
	\item the building up of the supply chain, during which the criteria determining the demands, prices, costs, production means and income sharing are defined, and
	\item the sharing of the income.
\end{inparaenum}
The production and marketing stages in between are out of scope for this paper. 
As outlined in Section~\ref{sec:background}, most of the efforts have indeed been devoted to the blockchain-based automation and management of those phases, whereas we focus on the use of blockchains as a means to optimise the selection and mutual benefit of the involved counterparts.
Notice that the beginning stage is key as the decisions taken therein (the selection of the core criteria and subsequent assembly of the chain) have a reentrant effect on the ex-post phase (as the income sharing scheme was decided at the beginning). 

In the remainder of this section, we introduce a motivating scenario and provide a high-level view of the income sharing approach and of the special role of service providers.

\subsection{A motivating scenario}\label{sub:scenario}
In many production environments, behind every product and service there are several companies dedicated to the different activities which, together, are necessary to translate primary resources into a product or service.
This is, indeed, the essence of supply chains. 
In a productive sector comprised of various  SMEs, such as the agri-food area, 
the contribution of the individual companies is fundamental.
The operational capacity of the 
%
%
chains in which they participate
%
%
%
requires a higher level of complexity as problems and decisions request the involvement of the consortium since they may not be solved by the single company. 
Therefore, the problems of chain management and integration become fundamental to make them more and more efficient~\cite{DBLP:books/sp/DumasRMR18}.

Some problems of the agri-food chains are the significant rise in prices from primary production and the excessive level of intermediation.
These issues are tightly connected.
One can typically observe a lack both of ``vertical'' and ``horizontal'' coordination, whereby the former refers to the relationship between supplier and customer and the latter to the cooperation between 
suppliers of the same resource.
However, the low level of integration is, in turn, caused by the difficulty of finding a company that is able to effectively and authoritatively cover the role of central coordinator among the SMEs in the chain~\cite{TA02a}.
By ``authoritative'' we mean the ability and even the possibility of exerting conditions on the behaviour of all the involved companies so that they act in the general interest of the entire chain considered as a whole. 
The system we present, on which the IS algorithm is based, is aimed at providing an answer to this problem.

The LEMURE project (MUR 4451/ICT) implemented and applied  the Revenue Sharing algorithm  (part of the IS algorithm as presented next in Section~\ref{sec:algorithm}), in relation to an agri-food chain for products ready for consumption.
The supply chain included a partner dedicated to the transformation and processing phase of a series of raw materials (tomatoes, oil, spices) coming from multiple suppliers. The production chain included also elements for packaging 
coming from other suppliers. The final products were distributed on the market at a price 
computed in response to the demand. 

Within the entire decentralised coordination approach~\cite{Tononi2007LaSC}, the production chain was defined by aggregating 
the fundamental costs of the production process 
(based on variable costs, fixed costs, production, time of recurrence, and the bill of material).
It was decided, also in relation to the supply chain optimisation approach and the constrained linear programming, 
The solution to optimise the supply chain was based on constrained linear programming.
Therefore, data were represented with a matrix structure so that the chosen formulations could be generalised regardless of the length or width of the supply chain under analysis.
Similarly, a number of parameters were used, functional to the definition of the production process as above 
%
%
alongside others that were defined in relation to the profit sharing phase 
%
%
(negotiated quota, price, demand of the final product).
We now proceed to discuss our conceptual setting in detail.
  
\subsection{Income sharing}\label{sub:sharing}
We recall here the characteristics of the IS algorithm, revising the presentation in~\cite{BGMP+20}.
The algorithm leverages a series of matrices associated with the supply levels that set up the chain description.

In this view, the supply chain is rooted in the request from an \emph{originator}, who advertises the need for a number of intermediate products and services in given quantities.
When a request is advertised, available and interested \emph{suppliers} (forming a set $M$) tender to provide the necessary products and services
(collectively called \emph{resources}, in a set $K$), also relying on other suppliers of resources that precede them in the process. 
This mechanism establishes a hierarchy of levels, ending with suppliers of raw materials or basic services that are self-sufficient to satisfy a request.
%
%
%
We denote levels with an integer $i\in{I}=[0,n]\subset\mathbb{N}$. 
In a multi-level structure (with $n\geq{2}$), resources to be provided to the higher level ($i{+}1$) may need other resources from the lower level ($i{-}1$).
In all supply chains, we place the market 
%
%
at level 1 and descend the levels in the hierarchy (i.e., follow the supply chain) as far as needed.
%

Notice that a supplier could operate at different levels supplying different resources: for example, in a cheese production process
a farmer could provide both fodder for animals to a breeder and vegetable rennet for cheese production.
The structure at the basis of the supply chain is thus represented as a relation $\SCS\subseteq{I}\times{K}\times{M}$. 
With reference to the previous example, the farmer will be involved in the chain at two levels, each provision being represented by a specific node in the chain. 
The farmer will then receive quotas according to the type of resource and the quantity supplied. 
%
Indeed, as discussed in~\cite{Tononi2007LaSC}, the construction of the chain 
%
%
is driven by the product resource rather than by supplier identity. 

We assume $K$ and $M$ to be finite non-empty sets and, as such, we can map every $k\in{K}$ and every $m\in{M}$ to an integer in the intervals $[0,|K|{-}1]$ and $[0,|M|{-}1]$, respectively.
%
%
In other words, each item in 
%
%
{\SCS} can be identified with the triple of indices $(i,k,m)$, where $i$ corresponds to the level, $k$ the resource type in a specific level and $m$ the supplier of the resource $k$ at level $i$. 
We shall also use the $(i,k)$ pair to identify the provision of resource type $k$ at level $i$, that is, such that there exists 
$m\in{M}$ such that $(i,k,m)\in\SCS$.

In order to create a request, the originator provides the following parameters:

\begin{description}[topsep=2pt,itemsep=2pt,parsep=2pt,partopsep=2pt]
\item[\normalfont{$\demand\in\mathbb{N}$}]: demand of the final product from the market;
\item[\normalfont{$\price\in\mathbb{R}^+$}]: price;
\item[\normalfont{$\BOM{i,k} \in \mathbb{R}^+$}]: the \emph{Bill of Material}, maping each type of resource $k\in{K}$ in the supply chain at each level $i\in{I}$ in which $k$ appears to a quantity --  that is, the ratio of the contribution of $k$ to the final product.
\end{description}
The originator can set upper bounds to the number of types of resources to be used ($\resmax \in \mathbb{N}$, so that $k\leq\resmax$), 
%
%
the number of levels ($\lvlmax \in \mathbb{N}$, hence $i\leq\lvlmax$), 
%
%
and the number of participating suppliers 
($\splmax \in \mathbb{N}$, with $m\leq\splmax$).

A supplier $m$
%
%
contributing a resource of type $k$ 
(at a certain level $i$)
must in turn characterise its contribution with various parameters:
\begin{inparadesc}
\item[\normalfont{$\fixc{i,k,m} \in \mathbb{R}^+$:}] fixed production cost;
\item[\normalfont{$\varc{i,k,m} \in \mathbb{R}^+$:}] variable production cost;
\item[\normalfont{$\qty{i,k,m} \in \mathbb{R}^+$:}] quantity of provided resource;
\item[\normalfont{$\tspan{i,k,m} \in \mathbb{R}^+$:}] time span to cover cost (in days);
\item[\normalfont{$\iquota{i,k} \in \mathbb{R}^+$:}] income quota (negotiated with other suppliers of $k$ at level $i$).
\end{inparadesc}
  
\subsection{Services and investors}\label{sub:servicesAndInvestors}
In the previous paragraph, we have outlined the input data provided by a supplier, which can be generally summarised in the production costs of the partners in the production chain.
In the IS algorithm, partners join forces to maximise profit.
The entry of a DAO into the SCM leads to different business scenarios and approaches in the DeFi field.
In fact, let us consider a new subject, configured either as a supplier of services for the supply chain, or as an investor that believes in the added value that the DAO provides to the decreed cooperation model.
What does this mean in terms of the originator's configuration of the supply chain, in terms of costs, or in terms of participation in the sharing profits?
Following the 
%
%
scenarios detailed in Section~\ref{sec:background}, the coordination via IT platform leads to considering three scenarios in the income sharing model:
\begin{enumerate}[topsep=2pt,itemsep=2pt,parsep=2pt,partopsep=2pt,label=(\arabic*)]
    \item the presence of a new subject in the chain (the IT platform provider) that shares both the risk and the final profit of the chain ($\PC$),
	\item the charging of costs of the IT services to the originator, or
	\item the division of costs of the IT services among all partners.
\end{enumerate}

In case the platform acts as an external service provider, hence not involved in the production aspect of the chain, (scenarios (2) and (3)), the related costs incur for the members of the chain to align themselves.
These 
%
%
alignment costs
%
%
are then computed and indicated here as \alcost. 

Note that in the chain model the aim of the alignment matrix is threefold:
\begin{enumerate}[topsep=2pt,itemsep=2pt,parsep=2pt,partopsep=2pt,label=(\roman*)]
    \item alignment of quality characteristics;
    \item product innovation;
    \item process innovation (our case).
\end{enumerate}
When the alignment cost occurs, then the $m$ 
member will receive, from thew overall proceeds, its compensation as a recovery for these costs.

\section{A modular perspective on contracts}\label{sec:modular}
We propose a software infrastructure enabling a platform to propose to its users 
(i.e., originators of and participants in a consortium) different options for income sharing and for the preferred form of financing 
(i.e., having the platform itself, the originator, or some specific member act as finance provider).

\begin{figure}[htb]
    \centering
    \includegraphics[width=12cm]{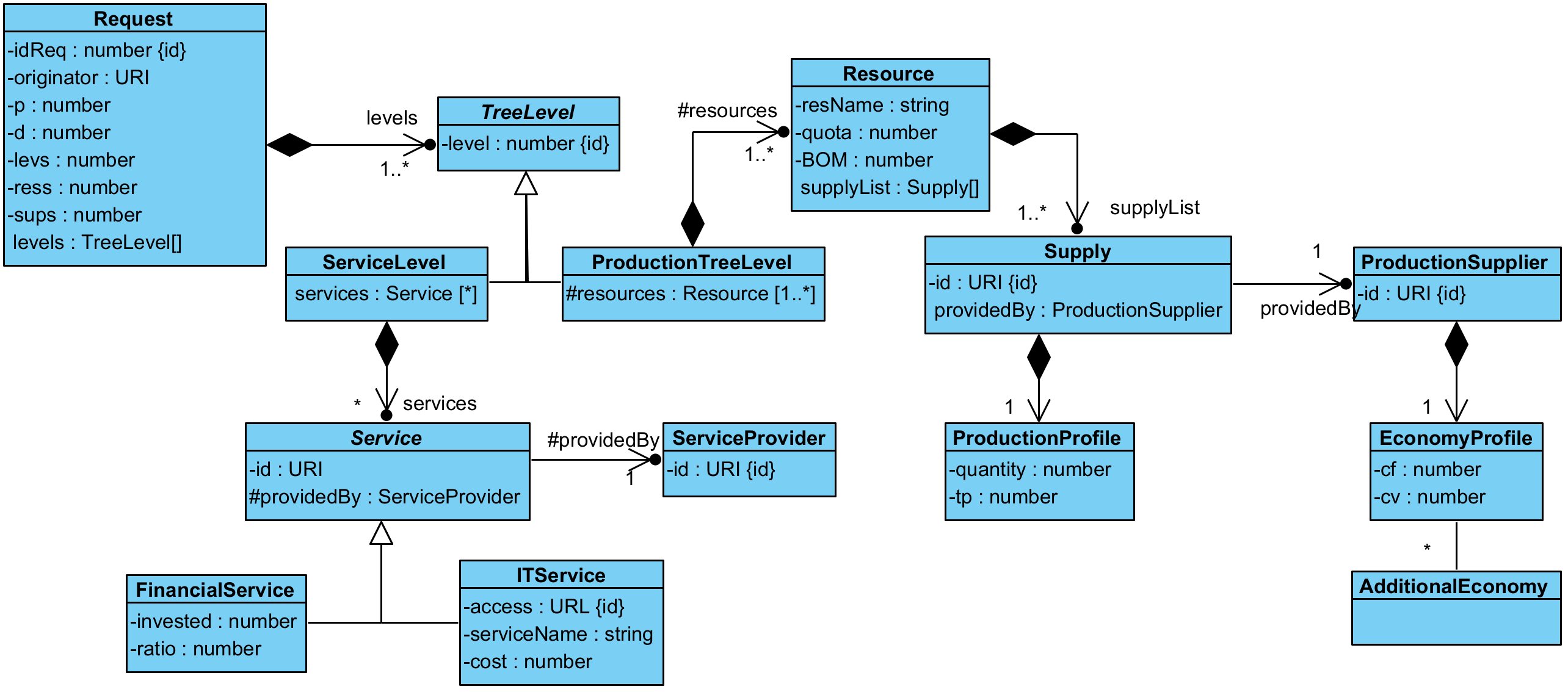}
    \caption{The conceptual model for the data structure supporting income sharing.}\label{fig:conceptualModel}
\end{figure}

The proposal is based on the conceptual model in Figure~\ref{fig:conceptualModel}, while concrete data structures and aspects of its implementation  
%
%
(for the Hyperledger Fabric framework) are presented in Section~\ref{sec:interacting}.

The model supports the separation of concerns between the production organisation of the supply chain structured according to the notions of \texttt{Resource}, \texttt{Supply}, and \texttt{ProductionProfile}, and the \texttt{EconomicProfile} pertaining to each specific \texttt{ProductionSupplier} for a given \texttt{Supply}. 

A supply chain is uniquely associated with the \texttt{Request} put forward by an \textit{originator}, and is structured as a collection of \texttt{TreeLevel}s. 
In particular, a request is described by attributes corresponding to the parameters discussed in Section~\ref{sec:supplyChain} (i.e., $p$ for price, $d$ for demand, etc.). 
Each \texttt{ProductionTreeLevel} describes the resources needed to realise a semi-finished product to be used at the level above, up to the finished product to be sold on the market. Each type of resource is characterised by a \textit{name}, a \textit{quota} of income to be reserved for the overall production of that resource and a list detailing how it will be supplied.

We assume that each supply chain produces just one type of product 
(possibly a packaging of a number of products) towards the market, this product being seen as described by the request, conceptually thought to be present at level 0, thus constituting the root for the tree representation of the supply chain.

The final level is complemented by the information at the only instance of \texttt{ServiceLevel}, describing \texttt{Service}s of different types, each provided by a \texttt{ServiceProvider}.
We currently envisage that the platform can give access to \texttt{FinancialServices}, the provider of which is an investor expecting the \textit{investment} amount to be remunerated at an income \textit{ratio}.
Other types of service are generic \texttt{ITService}s, of which the platform itself, identified by a \textit{serviceName}, \textit{access}ed through a URI, and exposing a \textit{cost} towards the consortium.

Since the model is agnostic with respect to the implementation, we use the dummy type \textit{number} to indicate that the value is a numerical one. 
This corresponds to the type \textit{number} in the Ethereum implementation (see~\cite{BGMP+20}), or to floats or integers in the TypeScript/JSON implementation (see~\cite{DBLP:conf/iscc/BottoniPTGM21}).

The model of Fig.~\ref{fig:conceptualModel} allows for a flexible and extensible realisation of consortium supply chains, where the choice of the Profit Sharing or Revenue Sharing scheme is orthogonal to the definition of the productive structure. 
The economy descriptors in the \texttt{EconomyProfile} can be used in both schemes to evaluate costs, while the actual remuneration depends on the chosen scheme.
The same will apply for IT Services, where costs are considered to be alignment cost for the whole consortium. 
Investors will be remunerated according to the requested ratio on their investment (the latter also seen as an alignment cost).

Therefore, the actual computation of the amount to be assigned to each participant will consider the whole structure of costs, so that the difference between the two currently supported schemes is reduced to decreasing the income by costs or not. 
It is immediate to notice that this calls for a realisation of the Strategy pattern~\cite{Gamma.etal/1994:DesignPatterns}, so that the inclusion of a different scheme would be reduced to providing a different strategy for the scheme and, if needed, descriptors in the \texttt{AdditionalEconomy} structure, to be used in that scheme.

\section{Calculating sharing quotas}\label{sec:algorithm}
Based on the structure in Section~\ref{sub:sharing},
the algorithm to calculate the fair configuration of income sharing 
(once the final price is realised on the market) proceeds in five steps.

\noindent\textbf{Step 1}: For 
%
%
$(i,k,m) \in \SCS$, the associated cost $\cost{i,k,m}$ is computed as:
%

\begin{equation}
    \cost{i,k,m}= \frac{\fixc{i,k,m}}{\tspan{i,k,m} \cdot \qty{i,k,m}} + \varc{i,k,m}.
\end{equation} 
\noindent\textbf{Step 2}: 
The value of $\mincost{i,k}$ is calculated as the minimum advertised cost $\cost{i,k,m}$ among all suppliers for each resource $k$ at layer $i$.

\begin{equation}
    \mincost{i,k}= \operatorname*{min}\limits_{m \in \{m \in M: (i,k,m) \in \SCS\}} \left(\cost{i,k,m}\right).
\end{equation} 

\noindent\textbf{Step 3}: 
$\GI$, the \emph{gross income}, is computed on the basis of the price ($\price$) and quantity of products established by the originator.
Representing with $m$ the request advertised by the originator, we have
\begin{equation}
	\GI=\price\cdot\sum_{m \in M}{\qty{0,0,m}}.
\end{equation}

\noindent
The share $\ishare{k}$ of revenues to assign to the group of suppliers for resource $k$ and the total income at level $i$, {\itoti{i}}, are computed as follows:
\begin{equation}
    {\itoti{i}}=\sum_{k \in K} \ishare{i,k}\mbox{, with }\ishare{i,k}= \demand \cdot \mincost{i,k} \cdot \BOM{i,k}.
\end{equation}
The overall total income, $\itot$, follows:
\begin{equation}
    {\itot}=\sum_{i \in I} \itoti{i}.
\end{equation}

The share of revenues, $\ishare{i,k,m}$, is calculated for each supplier $m$ of the resource $k$ at layer $i$ using a coefficient $f_{i,k,m}$, reflecting the contribution of quantity $\qty{i,k,m}$ with respect to the sum of contributions for resource $k$: 
%
\begin{equation}
    \ishare{i,k,m} = \ishare{i,k} \cdot f_{i,k,m} \mbox{, with } f_{i,k,m} = \frac{q_{i,k,m}}{\sum\limits_{m \in M} \qty{i,k,m}}.
\end{equation}

\noindent\textbf{Step 4}: The overall profit of the supply chain is calculated keeping into consideration so-called 
\emph{alignment costs}, i.e., the extra-costs that some suppliers of the resource $k$ might have incurred with respect to the minimum cost, \mincost{i,k}. 
These are typically related to costs required by strongly innovative production processes, or to investments required to keep up with the expected quality standards for the supplied resource.
To this end, parameter $\scost{i,k,m}$ equates to the minimum advertised cost if the advertised one is higher than that, or $0$ otherwise 
(this deters suppliers from exposing costs due to inefficiencies of their processes).
The individual ($\alcosti{i,k,m}$) and total ($\alcost$) alignment costs are calculated as follows:

\begin{align}
    \alcost = {\sum_{(i,k,m) \in \SCS}{\alcosti{i,k,m}}}\mbox{, with } & \alcosti{i,k,m}{=} \frac{\ishare{i,k,m}}{\mincost{i,k}}(\cost{i,k,m}{-}\scost{i,k,m}) \\
    \textrm{and } & \scost{i,k,m} =
        \begin{cases}
            \mincost{i,k,m} & \textrm{if } \cost{i,k,m} > \mincost{i,k,m} \\
                          0 & \textrm{otherwise.}
        \end{cases} \nonumber
\end{align}

We can now subtract these costs from the total income, before proceeding to calculate the sharing of the net income. 
Hence, denoting with $\PC$ the net \emph{profit chain}, with $\GI$ and {\itot} calculated at step 3,
and {\alcost} as above, we have 
\begin{equation}
	\PC = \GI - \itot - \alcost.
\end{equation}

\noindent\textbf{Step 5}: For each participant, the $\GIs{i,k,m}$ share is
%
%
established
based on the
%
%
quotas of resources provided by each supplier as per some previous negotiation,
%
%
%
with respect to each resource $k$ based on the negotiated requirements $\iquota{i,k}$, under the constraint that
\[ \sum_{\substack{i \in I\\ k \in K}}{\iquota{i,k}}=1. \textrm{\quad }\] 
%
We denote with $G$ the matrix consisting of $\iquota{i,k}$ elements for every $i \in I, k \in K$.

Then, each partner receives a specific revenue profit \GIs{i,k,m}, calculated considering the income quota \iquota{i,k} (negotiated for every resource) and the specific supplied quantity \coeff{i,k,m}. 
To this end, we use the following formula:

\begin{equation}
   \GIs{i,k,m} = \coeff{i,k,m} \cdot \iquota{i,k} \cdot \PC.
\end{equation}
%
%
This last step closes the loop and highlights the fundamental difference between IS and vertically integrated chains dominated by subjects who, being financially much stronger than the others, can impose downward auction mechanisms with effects of economic starvation of the weaker suppliers. 
In fact, on the one hand an IS consortium results from the lighting up by an originator of the entrepreneurial fuse of the initiative, yet on the other hand even those companies that join subsequently will be entitled to fair compensation through the distribution of the proceeds. Furthermore, the lower the internal supply costs are kept, the larger the returns will be as an effect of greater competitiveness on the market. 
Therefore, we can say that IS provides the effectiveness of vertical integration yet maintains an equitable distribution of returns, according to a mechanism that can boost the competitiveness of consortia of small and medium-sized enterprises against giants such as Amazon and Walmart, as extensively argued and illustrated in~\cite{BGMP+20}. Also notice that the algorithm only comes into play for the purpose of managing freely negotiated agreements, as set out in the process of definition of costs and prices above.

Following the scenarios detailed in Section~\ref{sec:supplyChain}, we consider now the case in which the IT platform is provided by a external service provider. 
Then, the alignment cost can be borne by the originator or divided among all the partners.
%
%
This means that 
%
%
%
  {$\cost{i,k,m}>\mincost{i,k}$}
and the 
%
%
$m$ member should receive the following compensation 
%
%
for 
the alignment costs:  
\begin{equation}
    \alcosti{i,k,m}  = \frac{\ishare{i,k,m}}{\mincost{i,k}}\cdot(\cost{i,k,m}{-}\scost{i,k,m}).
\end{equation}

At this stage, we can configure 
the $i$, $k$, and $m$ indexes depending on whether the originator or all the partners bear the costs. 
When only the originator provides this cost, assuming that the originator is the second provider ($m=2$) of resource $k=1$ at the first level of the chain ($i=1)$, 
we have:
 \begin{equation}
    \alcosti{1,1,2}{=}\frac{\ishare{1,1,2}}{\mincost{1,1}}(\cost{1,1,2}{-}\scost{1,1,2}).
\end{equation}
%
\noindent
This will be the only value different from $0$ in the $cAll$ matrix. \\
If instead all partners cover the cost of the IT platform provider, 
%
%
then Equation~(10) is applied for each node in the chain. 
If the IT platform provider is a member of  the chain, she takes part of the profit as described in~\cite{Gong2018RevenueSO} with a specific income quota. 
In the RS model presented in~\cite{DBLP:conf/iscc/BottoniPTGM21}, the $G$ matrix is used to express the cutting ratio of the different resource providers. 
%
%
Suppose now that 
a third-party IT provider joins the consortium.
As the 
%
%
matrix indexing scheme is flexible, it can manage new partners, whether in the production chain or outside it (as in this case with an IT platform provider). 
The IT platform provider is not product-oriented. 
Hence, following the approach described in~\cite{Tononi2007LaSC},
the most suitable solution is to add a final ``third party'' level under level $n$. 
This extra level acts only on the $G$ matrix, where the negotiated requirements are stored.
This means that only in $G$ do we find at level $n+1$ a value that is different from zero and we can manage this value according to the model in~\cite{Gong2018RevenueSO}.

\section{Interacting with the smart contracts}\label{sec:interacting}
We present here an overview of how the participants in a supply-chain consortium would interact with the platform, 
through the interfaces for \textit{originator}, \textit{supplier}, \textit{IT provider} and \textit{finance provider} (\textit{investor}).

As discussed before, we assume that the choice of the usage configuration of the platform has been made offline, during negotiations for the formation of the consortium.
Hence, the originator has two tasks: 
\begin{inparaenum}[(1)]
\item  to define the overall constraints (e.g., the number of levels) and the top level of the productive structure (i.e., the types of resources and relative quantities to obtain the end product); and 
\item to set the platform to operate under the agreed options (Profit or Revenue Sharing, presence of an investor, required IT services). \end{inparaenum}

\begin{figure}[htbp]
	\centering
	\includegraphics[width=\textwidth]{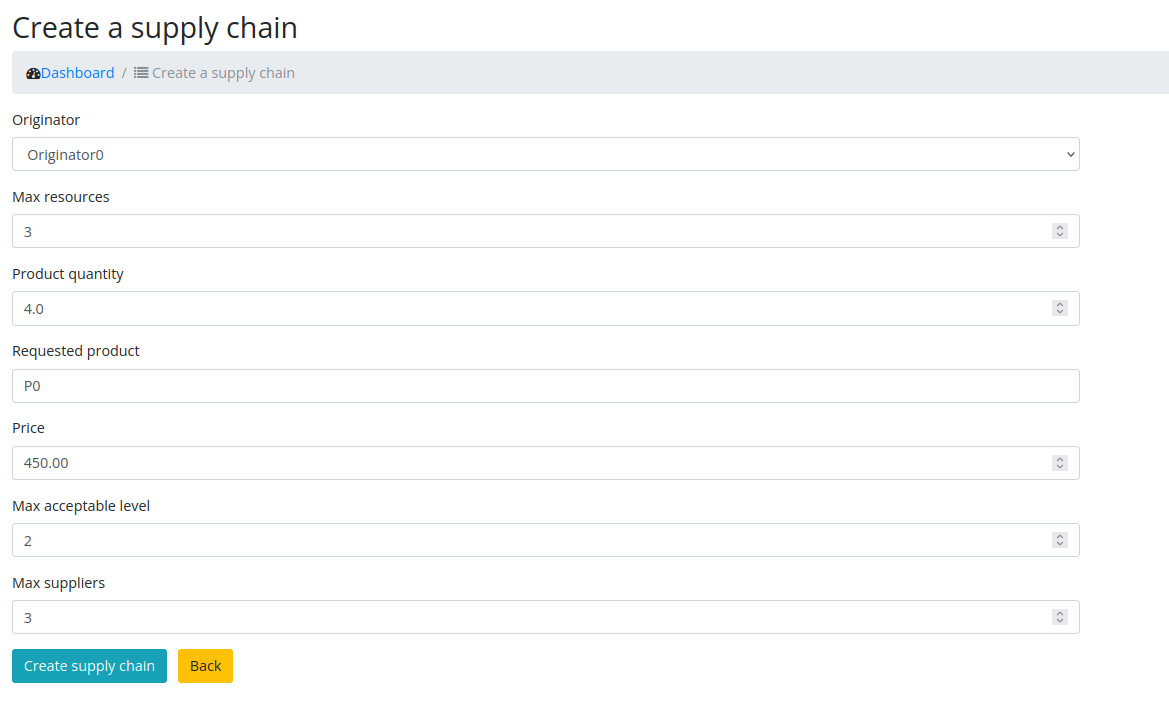}
	\caption{The form for the specification of the request properties}
	\label{fig:createRequest}
\end{figure}
\begin{figure}[htbp]
    \centering
    \includegraphics[width=\textwidth]{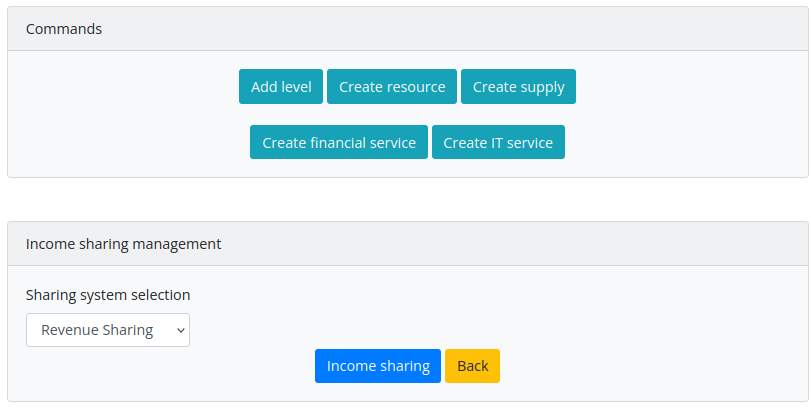}
    \caption{The dashboard for the configuration of the consortium structure}
    \label{fig:dashboard}
\end{figure}

Figure~\ref{fig:createRequest} presents a screenshot of the form for the specification of the properties characterising a request. The form is integrated with a dashboard to set the options related to the sharing scheme and the presence of external services (see Fig.~\ref{fig:dashboard}).
In particular, the latter lists the possible actions that a user can perform on the supply chain tree to add new elements to the structure. 
The last command executes the revenue sharing algorithm on the actual supply chain.

Once these options have been chosen, each participant in the system is presented with an interface to enter values for the parameters required by the selected configuration.
Typically, the interface provides a field for every descriptor in the production and economy profiles (possibly including the additional economy descriptors) to be set by the supplier.

\begin{figure}[htbp]
	\centering
	\includegraphics[width=\textwidth]{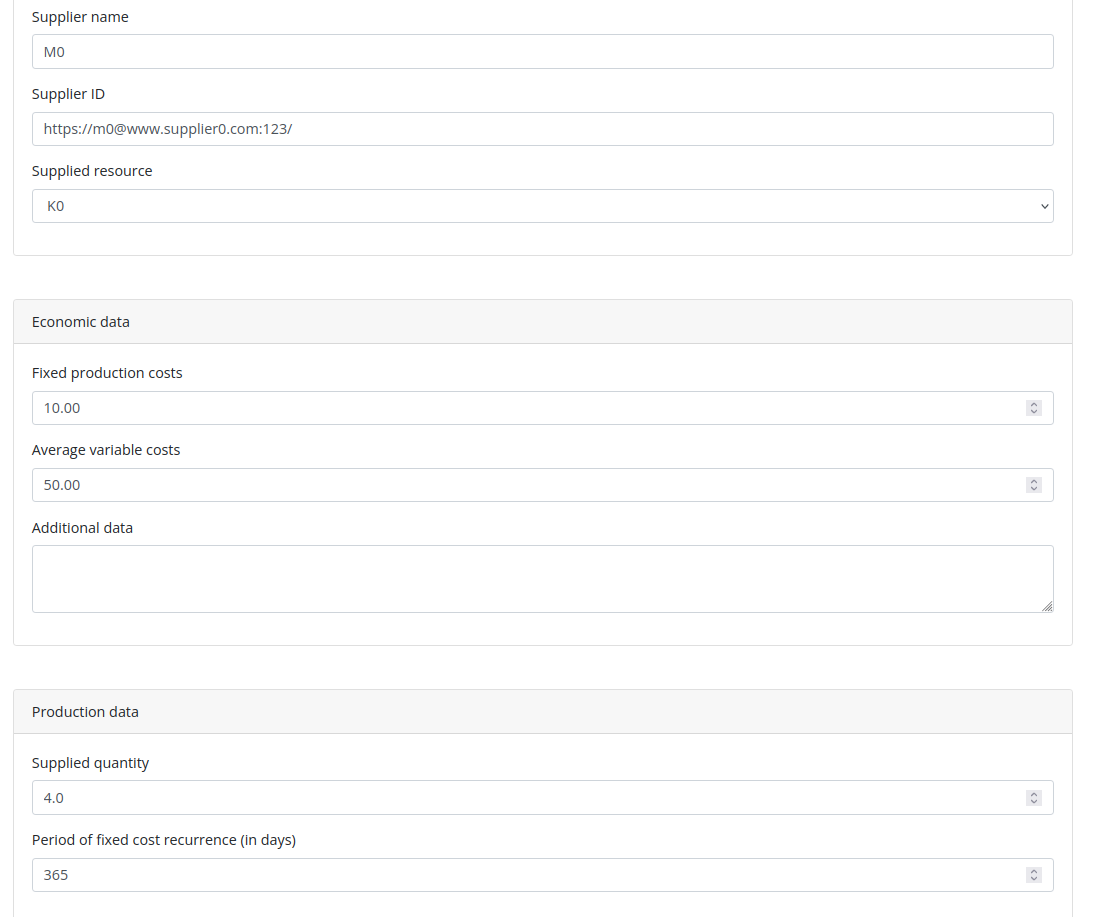}
	\caption{The interface for the product supplier.}
	\label{fig:productSupplierInterface}
\end{figure}

In particular, Fig.~\ref{fig:productSupplierInterface} depicts the interface for a supplier of a semi-finished product. It consists of a form divided in three main sections: in the first section, suppliers can insert data identifying their contribution, while the second one is for defining the economic profile related to that supply, declaring fixed and variable costs.  
The last section is meant to be filled with the production data (supplied quantity and period of cost recurrence).

\begin{figure}[htbp]
	\centering
	\includegraphics[width=\textwidth]{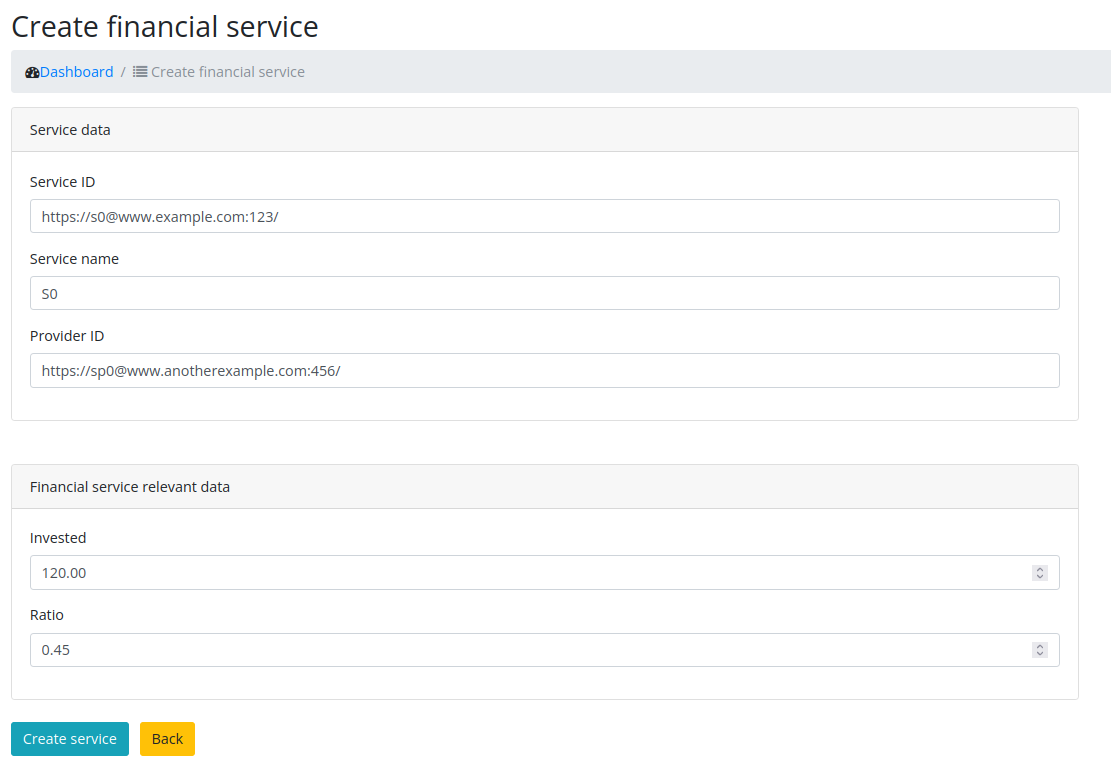}
	\caption{The interface for providers of financial services}
	\label{fig:financeProviderInterface}
\end{figure}

\begin{figure}[htbp]
	\centering
	\includegraphics[width=\textwidth]{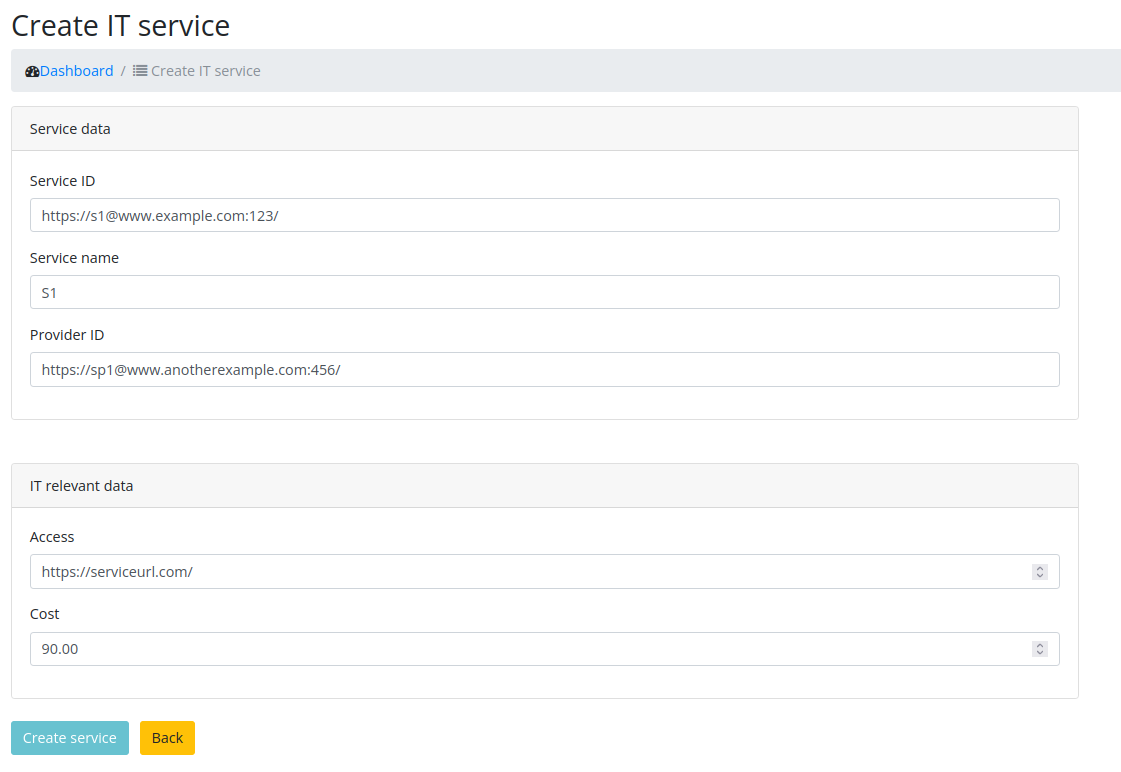}
	\caption{The interface for providers of IT services}
	\label{fig:serviceProviderInterface}
\end{figure}

Figures~\ref{fig:financeProviderInterface} 
and~\ref{fig:serviceProviderInterface} show the interfaces that allow the user to define, respectively, financial and IT services. 
Some descriptors are common to all types of service (e.g., \textit{name}, \textit{ID} of the service as a URL, and \textit{ID} of the provider as a URI).
In addition, financial services are defined by the \textit{invested} amount of money and the cutting \textit{ratio}, IT services by their \textit{cost} and the URL for \textit{access}.

When all the suppliers and providers have uploaded their information, the resulting supply chain structure can be stored in a form specified by the JSON scheme of Listing~\ref{lst:json}, (edited here for the sake of readability).
Each class is represented as a structure, scoped by a pair of curly brackets, while lists of elements with a given structure are surrounded by square brackets. 
Numerical types are considered to be implemented either as float associated with the default `0.0', or integers, for which the default is `0'. String types, including URIs and URLs, are set to empty string if not specified otherwise.

\begin{figure}[htbp]
\begin{minipage}{\linewidth} 
\begin{lstlisting}[language=json,caption=An excerpt of the JSON descriptor for the case at hand,label={lst:json}]
{
    "requestId": 1, "originator": "Originator0", "p": 450, "d": 4, "levs": 2, "ress": 3, "sups": 4, 
    "levels": [ {
        "i": 1, 
        "resources": [ {
            "resourceName": "K1", "g": 0.4, "BOM": 8.0,
            "supplyList": [ { "m": 0,   
                "supplierData": { "supplierName": "M0",
                    "supplierId": "https://m1@www.supplier1.com:456/" },
                 "economicProfile": { "cf": 35, "cv": 35, "additionalData": {} },
                 "productionProfile": { "q": 12, "tp": 365 } } ] }, (* \hfill\textit{\textrm{\ldots more resources listed here\ldots}} *)
        ] }, (* \hfill\textit{\textrm{\ldots level 2 details listed here.}} *)
    ],  
    "serviceLevel": {
        "financialServices": [ { "serviceName": "S0",
                    "uri": "https://s0@www.example.com:123/",
                    "providerId": "https://sp0@www.anotherexample.com:456/",
                    "invested": 120, "ratio": 0.45 } ],
        "itServices": [ { "serviceName": "S1",
                    "uri": "https://s1@www.example.com:123/",
                    "providerId": "https://sp1@www.anotherexample.com:456/",
                    "access": "http://www.serviceurl.com", "cost": 90 } ] } }
\end{lstlisting}
\end{minipage}
\end{figure}

%
%
%

\begin{figure}[htbp]
    \centering
    \includegraphics[width=0.8\textwidth]{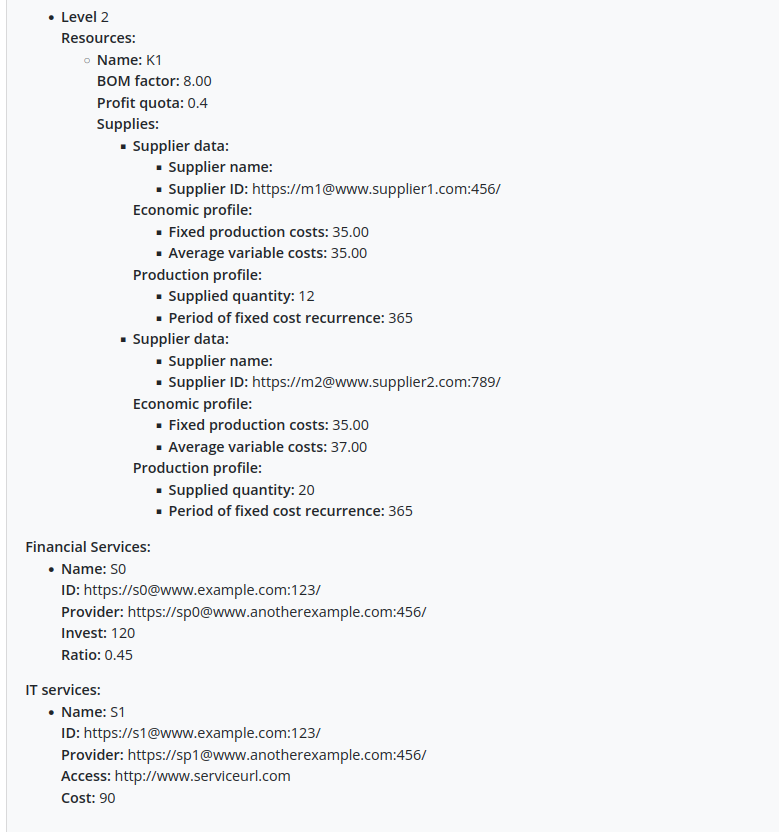}
    \caption{A summary of a configuration for a supply chain}
    \label{fig:details}
\end{figure}

%
%
As a visual counterpart, Fig.~\ref{fig:details} shows the page presented to participants to summarise the resulting configuration of the supply chain (request) data. 
The summary starts with the overall descriptors of a request, and then, for each level in the chain, presents a list of supplied resources. 
For each resource type, properties relative to the resource itself and the relative supplies are shown.
The first level presents the information about financial and IT services.
With each set period, given the reaped incomes in that lapse of time, the algorithm applies the strategy corresponding to the selected sharing scheme to evaluate the quotas for the different participants.

\begin{figure}[htbp]
	\centering
	\includegraphics[width=\textwidth]{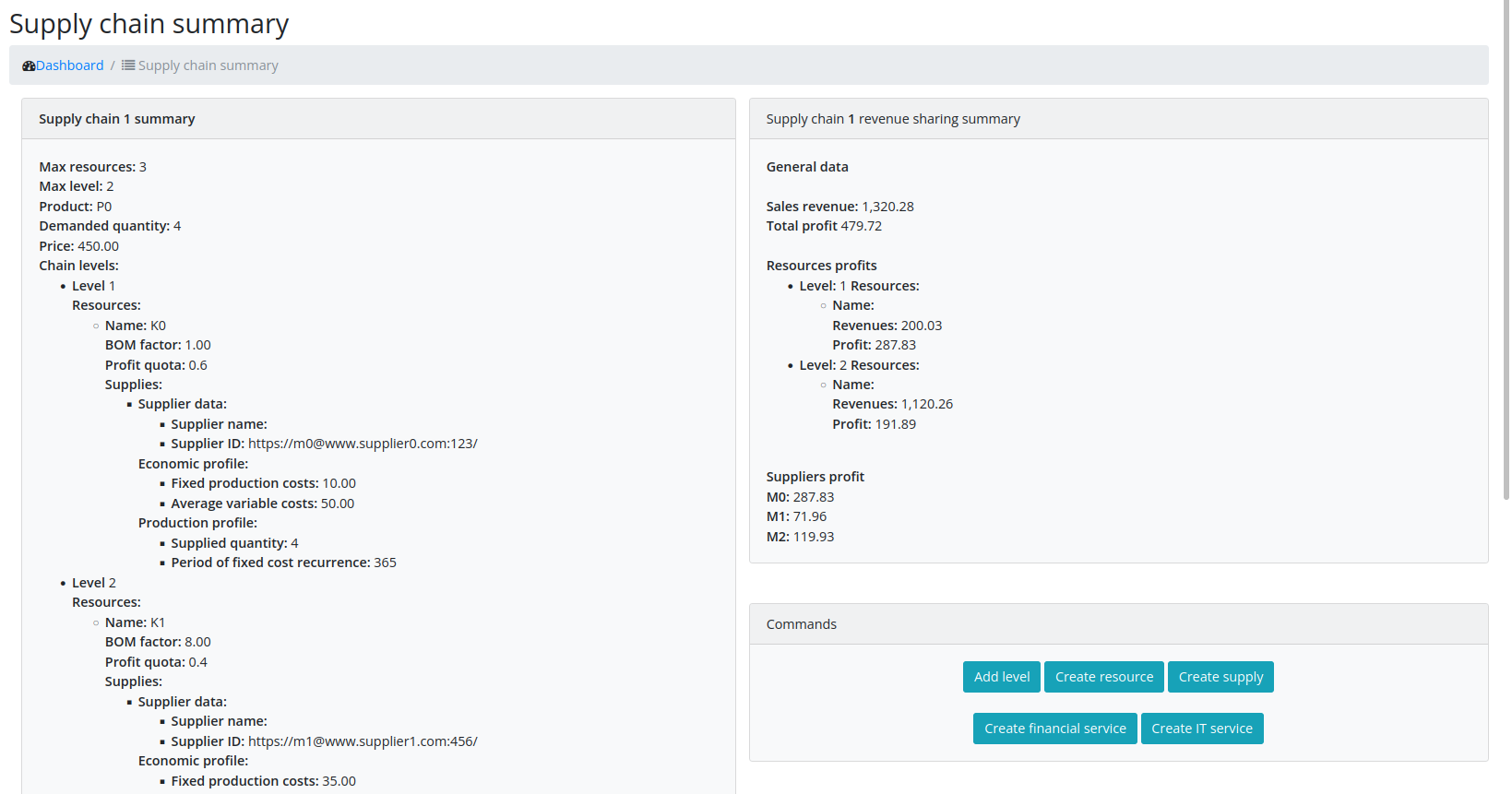}
	\caption{The interface showing the final sharing of proceeds}
	\label{fig:detailsWithRevenues}
\end{figure}

Finally, Fig.~\ref{fig:detailsWithRevenues} depicts a screenshot representing the situation after computing the income quotas for the different participants (in this case, two service providers: the financing actor and the product supplier).

\section{Implementation trade-offs}\label{sec:discussion}
%
The prototypes for RS in~\cite{BGMP+20} and \cite{DBLP:conf/iscc/BottoniPTGM21} have been implemented in Ethereum and Hyperledger, respectively. 
We have extended the latter to encompass the Income Sharing architecture.
The choice between the two options represents a crossroads from the viewpoint of implementation, each with its pros and cons as illustrated next. 
However, 
we remark that this divergence of directions can be recomposed through interchain integration mechanisms aiming to combine the best of the two possible worlds, namely public blockchains, of which Ethereum is the most prominent representative as for industrial applicability, and private or consortium DLs, 
%
%
similarly represented by the Hyperledger software ecosystem.

The aspects to be considered in order to make that choice essentially correspond to the trade-offs between \emph{algorithmic trust} and \emph{human trust}, on one hand,
and the balance among \emph{privacy}, \emph{computational efficiency} and \emph{consensus robustness}, on the other hand.
%
%
Ethereum maximises the algorithmic trust guaranteed by the thousands of nodes of its network and the robustness of its validation protocol based on 
Proof-of-Work (PoW). This is because Ethereum could be put to test by nodes with fraudulent intentions, which are not directly identifiable due to the pseudonimity guaranteed by the platform. 
For the same reason, by enabling global interaction between companies through algorithmic trust, it lends itself perfectly to an optimal implementation of the principles of Transaction Cost Economics.
Yet, all this exacts a high toll on efficiency as a consequence of the costs of block construction.
Also, it requires that method invocations are paid by means of gas units.
As the price of cryptocurrencies is subject to oscillations with an upward trend, this aspect could become far from negligible in the presence of numerous transactions.
Furthermore, since the blockchain is public, the privacy of transaction payloads is not guaranteed -- an aspect which can be a deterrent for the adoption of the system in an enterprise context. 
This provides a likely reason for the losing ground of Ethereum versus Hyperledger in the first wave (essentially focused on tracking production processes) of supply chain projects on blockchain and on distributed ledger technology, as reported in~\cite{VadgamaT21}.

By contrast, opting for Hyperledger moderates the role of algorithmic trust; yet, this aspect may be compensated by governance mechanisms where the various participants know each other. 
Privacy can be flexibly managed at various levels, starting from the permissioned nature of the participation. 
Computational efficiency is also increased by virtue of the smaller number of nodes and the computationally less demanding consensus protocols that can be applied assuming a higher level of human trust (such as the Crash-Fault Tolerant protocols adopted in Huperledger Fabric\footnote{\url{https://www.hyperledger.org/wp-content/uploads/2017/08/Hyperledger\_Arch\_WG\_Paper \_1\_Consensus.pdf}}). 
In any case, the trust management systems in place guarantee, despite having a more limited role than in public blockchains, a level of automation and reliability such as to avoid, in order to join the organisation, the long and expensive preliminary checks typical of pre-digital economy. Furthermore, the fact that governance is not completely alienated in favour of a totally algorithmic set-up facilitates corrective intervention in the face of emergencies and anomalies, an aspect that has been decisively lacking in the case of the DAO Exploit as pointed out by Morrison~\textit{et~al.}~\cite{DBLP:journals/fbloc/MorrisonMW20}. This feature, combined with the focus and goal-orientation of the Income Sharing consortia, appears particularly promising for their robust and effective functioning.

A further concern derives from being bound in Ethereum to using its dedicated language (Solidity) and virtual machine (EVM). 
By contrast, Hyperledger supports a large variety of very popular languages. In particular, Fabric, the most popular among the Hyperledger frameworks and the one adopted here for the implementation of the IS architecture, supports Java and JS, in addition to the more recent Go, and can be deployed through much more common and widespread installation setups (NPM, Node, Go, Docker). 
This obligation is not to be taken lightly: although optimised for EVM, Solidity is a young language, therefore still immature in some respects, which could hinder large scale developments. 
A specific limiting factor is  that a rather small memory stack is dedicated to declarable variables.
For IS algorithms that involve a large number of parameters, this makes it necessary to distribute the arguments over several functions that need to call each other, with consequent unnecessary complications in the structure of the code.
Another problematic aspect is that numeric variables in decimal format are poorly supported, which is of no help for the calculation of income shares, where operations with decimal numbers are the order of the day. 
Garriga~\textit{et~al.}~\cite{GarrigaPARPT21} and Palma~\textit{et~al.}~\cite{Zappone21} illustrate methodologies that are applicable to Income Sharing, among others, to choose on the basis of these and other trade-offs among the Hyperledger platforms (of which Fabric is the best known) and the available public blockchains.

\section{Conclusions and future work}\label{sec:concls}
We presented an architecture for virtual consortium organizations, but with members corresponding to companies rooted in the economy, based on the general principle of Income Sharing between participants in a supply chain. We have illustrated its implementability on a blockchain or distributed ledger infrastructure which, through the use of smart contracts, guarantees both its optimality and reliability. We have shown how this architecture makes available a wide menu of optimized executions of supply chains attributable to different Income Sharing options, in a spectrum of choices ranging from Revenue Sharing to Profit Sharing, and we have included in the scenario the role played by service providers such as Amazon and other Amazon-like Internet platforms, whose role is ever more relevant in the global economy. We have also indicated the relationship between this architecture and the well-known Decentralized Autonomous Organization (DAO), of which it takes up the basic vision, however strengthening it in terms of economic concreteness as well as computational feasibility.

The implementation of this architecture was carried out in Fabric, the most successful platform to date for private and consortium distributed ledgers, developed within the Hyperledger ecosystem. We have indicated the advantages from the point of view of privacy and efficiency of this implementation choice. However, we have also highlighted how this type of implementation implies a decrease in the level of algorithmic trust,
 guaranteed instead by public blockchains. 
A future development we intend to pursue is to recover high standards of algorithmic trust by integrating the deployment of Income Sharing on a distributed ledger with a public blockchain capable of giving a further validity stamp to the processes thereby executed. 
In this perspective, the bulk of the work is done within the ledger, thus maintaining the aforementioned advantages, but some of the process milestones (such as the finalisation of income distribution) can receive further and definitive validation through a public blockchain. 
This type of integration is in fact provided specifically with Ethereum for the Besu and Burrow frameworks from the Hyperledger software ecosystem. Fabric itself is amenable, albeit not so automatically, but without restrictions with respect to the public blockchain to be integrated with (as for example illustrated in \cite{CarliniCPPZ20} regarding an integration between Fabric and the Stellar blockchain). 
As often happens, it is thus a question of putting together the best of two possible worlds.




\section*{Acknowledgment}
The work was partially supported by Sapienza University of Rome with the “Consistency problems in distributed and concurrent
systems” research project. 
This research work was performed under the MoU between ENEA, the Department of Computer Science at Sapienza, and the University of Molise.
The work of Claudio Di Ciccio was partly supported by the Italian Ministry of University and Research (MUR) under grant ``Dipartimenti di eccellenza 2018-2022'' of the Department of Computer Science at Sapienza and by the Sapienza SPECTRA research project.

\bibliographystyle{elsarticle-num}
\bibliography{biblio}

\end{document}